\documentclass[9pt]{book}

\usepackage[dvips]{graphicx,color}
\usepackage{makeidx,uni,verse}

\makeauthorindex
\BookTitle{Astronomy and Astrophysics of Extreme Universe}
\CopyRight{\copyright 2007 by Universal Academy Press, Inc.}

\begin{document}

%\tableofcontents
\pagenumbering{arabic}

\chapter{The Pierre Auger Observatory: status, results and perspective}

\author{\raggedright \baselineskip=10pt%
{\bf Etienne Parizot,$^{1}$ for the Auger Collaboration$^2$}\\ %%%%%% <== Authors
{\small \it %
(1) APC, Universit\'e Paris 7 Denis Diderot, 10, rue Alice Domon et L\'eonie Duquet, 75205 Paris, France\\
(2) Pierre Auger Observatory, av. San Martin Norte 304 (5613), Malarg\"ue, Argentina\\
}
}

\AuthorContents{E.\ Parizot, for the Auger Collaboration}
\AuthorIndex{Parizot}{E.}
\AuthorIndex{Auger Collaboration}{P.}
\baselineskip=10pt
\parindent=10pt

\section*{Abstract}

While the completion of the Pierre Auger Observatory (or simply ``Auger'') is still underway, the 5165~km$^2$~sr~yr integrated acceptance accumulated since the January 1st, 2004 is now significantly larger than what was gathered by the previous experiments dedicated to the detection of ultra-high-energy cosmic rays (UHECRs). We report on the development status of Auger and present some results related to the cosmic-ray energy spectrum, composition and anisotropies, and the photon fraction at ultra-high energy. We briefly discuss the importance of the ankle region to understand the overall phenomenology of cosmic-rays, and mention future enhancements of Auger focusing on this energy range.

\section{Introduction}

Cosmic-rays are energetic, charged particles traveling across the Galaxy and the whole universe under the influence of magnetic fields, which deflect their trajectory, and various nuclear processes resulting in energy loss, nuclear transmutation and the creation of secondary particles, including electron-positron pairs, pions, neutrinos and gamma rays. The energy spectrum of these cosmic rays (CRs) is a steep, roughly regular power law of logarithmic index $\alpha \simeq 2.7$--3, extending from thermal energies in the interstellar medium up to at least a few $10^{20}$~eV. The secondary particles produced by the interaction of cosmic rays in the Earth atmosphere have been known for a century (they actually led to the discovery of cosmic rays by Hess in 1912). They can also be used to detect energetic CRs, above a few $10^{14}$~eV, through two independent detection techniques exploiting the coherence of the so-called ``extensive air showers'' (EAS), produced by cascade reactions following the first interaction of a CR with an atom in the upper atmosphere: i) surface detectors (SD) sample the particles of the shower reaching the ground and analyze their lateral distribution (number density in the shower plane, perpendicular to the shower axis; ii) fluorescence detectors (FD) measure the fluorescence light induced in the atmosphere by the ionizing particles of the shower, as it develops from just a few very energetic particles high in the atmosphere to a maximum ionization power from many (billions or more) low energy particles at a given depth in the atmosphere, known as $X_{\mathrm{max}}$, measured in g/cm$^2$.

\section{Development status of the Pierre Auger Observatory}

The Pierre Auger Observatory is the first large aperture UHECR detector making extensive use of both detection techniques, SD and FD. It is located in the province of Mendoza, Argentina, and will cover a surface area of 3000~km$^2$ at an altitude of $\sim 1400$m. Completion is expected around the end of year 2007. The SD consists of an array of 1600 water tanks deployed on a hexagonal grid with a spacing of 1.5~km. These tanks detect the Cherenkov light produced by shower particles crossing their (1.2~m)$\times$(10~m$^2$) water volume, thanks to three 9-inch photo-multipliers (PMT). The geometry of a given cosmic-ray shower can be reconstructed from the arrival time of the shower front on the triggered stations and from the respective intensity of the detected signals. Given the Auger SD configuration, the interpolated signal 1000~m away from the shower axis can be inferred with satisfactory precision ($\sim 4$\%) for any shower energetic enough to trigger 3 (non-aligned) or more stations. This so-called $S_{1000}$ signal can then be related to the energy of the incoming cosmic ray either by comparing to Monte-Carlo simulations of EAS development (relying on the extrapolation of hadronic models constrained at lower energy by accelerator physics) or by calibrating this signal with the fluorescence signal measured simultaneously by the FD in the case when the shower is seen by both detectors, which is referred to as a \emph{hybrid event}.

The FD consists of four ensembles of six telescopes, each of which has a field of view of 30$^\circ$ vertically and 30$^\circ$ horizontally (i.e.~180$^\circ$ for each FD site). The telescopes are based on Schmidt optics and provide images of any (powerful enough) shower developing in the atmosphere above the SD array. Each telescope consists of: i) a filter at the entrance window, with very high efficiency in the 300--400~nm range, hosting the main molecular lines of Nitrogen, ii) an optimized circular aperture, iii) a corrector ring reducing spherical aberrations and keeping the spot size on the camera within 15~mm (which corresponds to an angular resolution of 5$^\circ$, i.e.~one third of the field of view of a single pixel-PMT), iv) a segmented $3.6\times 3.6\,\mathrm{m}^2$ mirror (with a radius of curvature of 3.4~m), and v) a camera made of 440 1.5-inch PMTs arranged in a $22\times 20$ matrix.

The FD gives access to two important parameters of an extensive air shower: the ``shower maximum'', $X_{\mathrm{max}}$, at a given energy, which depends on the mass of the incoming high-energy nucleus, and the total ionization power, which is directly related to its energy. This relation involves three important steps: i) the generation of the fluorescence itself, which depends on the \emph{fluorescence yield} in the atmosphere, ii) the transmission of light through the atmosphere (involving both absorption and diffusion), which is experimentally controlled thanks to intense atmospheric monitoring, and iii) the response of the cameras, which are calibrated in a relative manner at least twice a night and in an absolute manner less frequently, using a calibrated LED with known spectrum, intensity and directionality.

As of May 2007, all four FD buildings are operational and all 24 fluorescence telescopes are taking data. The deployment of the SD tanks is back to normal (after some non-technical issues related to land access permission), and 1200 stations, i.e.~3/4th of the full array, are operational and sending data, with an overall up time larger than 98\%. The SD and thus the whole Pierre Auger Observatory will be completed towards the end of 2007.

\section{Performance and energy reconstruction}

The key problem in high-energy cosmic ray experiments is the reconstruction of the shower energy. 

Identifying showers themselves is usually straightforward, as there is essentially no ``background'' for the detectors, at least above their energy threshold. In the case of Auger, the threshold for the SD is around 0.5~EeV, below which less than 10\% of the showers can trigger three tanks or more, as required. However, full detection efficiency (i.e.~100\%, or ``saturated acceptance'') is achieved only around 3~EeV for showers with zenith angle lower than 60$^\circ$, and lower energy showers are usually discarded to avoid any complication caused by the energy dependence of both the detection efficiency and the energy resolution. For the FD, showers with energies as low as 0.1~EeV can be observed. However, the corresponding acceptance is relatively low, since the total intensity of the fluorescence light does not allow detection from a large distance, and the shower maximum is then usually above the field of view of the telescopes, which prevents accurate reconstruction. Like for any fluorescence detector, the Auger FD acceptance increases with energy (as bigger showers can be seen from larger distances) and depends on the atmospheric conditions. However, a precise determination of the FD acceptance is not crucial for Auger, thanks to its hybrid nature, since the energy differential flux (or ``spectrum'') is not obtained from the FD, but from the SD whose absolute acceptance is essentially geometrical above saturation ($\sim 3$~EeV) and is thus controlled within a few percent at most.

%Since the FD can only work at night, with no or very little moon, only about 10\% of the useful Auger events are hybrid. The statistical power is thus obtained from the SD, while the FD is most useful for cross-calibration purposes as well as for an independent measurement of the $X_{\max}$ parameter, whose average and fluctuations at a given energy are directly related to the cosmic-ray composition.

The axis of development of the showers (indicating the arrival direction of the cosmic rays) is reconstructed with the SD by triangulation, using a GPS time tagging of the shower front arrival in the triggered tanks. With the FD, a fit of the track observed on the pixelized camera gives the plane containing the shower axis and the telescope, within which the axis itself is obtained thanks to the time information of each pixel, with improved precision when a signal from an SD tank is also available, i.e.~for hybrid events. The resulting angular resolution for the SD alone (i.e.~for most of the events) is better than 2$^\circ$ in the worse case of vertical showers with only three tanks triggered, and significantly improves for higher multiplicities, down to less than 1$^\circ$ for 6-fold (or more) events, i.e.~above $\sim 10$~EeV\cite{Ave_ICRC}. This is less than the expected deflection of the CR trajectories in the Galactic magnetic field alone (except for photons and neutrinos, of course), even at the highest energies, and angular accuracy is thus not perceived as a limiting factor in Auger, at least for the current analyses.

The energy reconstruction is more delicate, since the CR spectrum is a steeply decreasing function of energy and a misunderstanding of i) the link between the measured quantities (i.e.~secondary observables) and the actual incoming energy and/or ii) the underlying energy resolution (especially its dependence with energy) can have an impact on the reconstructed spectrum. Moreover, the threshold for the creation of pions and e$^+$e$^-$ pairs through the interaction of UHECRs with the CMB photons provides an absolute scale in this energy range. Testing the various astrophysical models against the reconstructed energy spectrum will thus be all the more efficient and meaningful that the absolute scaling of the UHECR data is more accurate. The hybrid design of the Pierre Auger Observatory is very useful in this respect. On the one hand, the FD measurements can provide a calorimetric estimate of the energy of cosmic-ray showers, while the traditional method to reconstruct the shower energy from the SD data involves a comparison with Monte-Carlo simulations based on extrapolations of the hadronic models beyond the energy range investigated in particle accelerators on Earth. On the other hand, the SD can be used to gather a large statistics with a well-controlled (geometrical) acceptance, while the FD detector is limited by its 10\% duty cycle (since it can only work at night, with no or very little moon) and an energy-dependent acceptance and energy resolution.

\begin{figure}[ht]
\begin{center}
\includegraphics[width=0.5\textwidth]{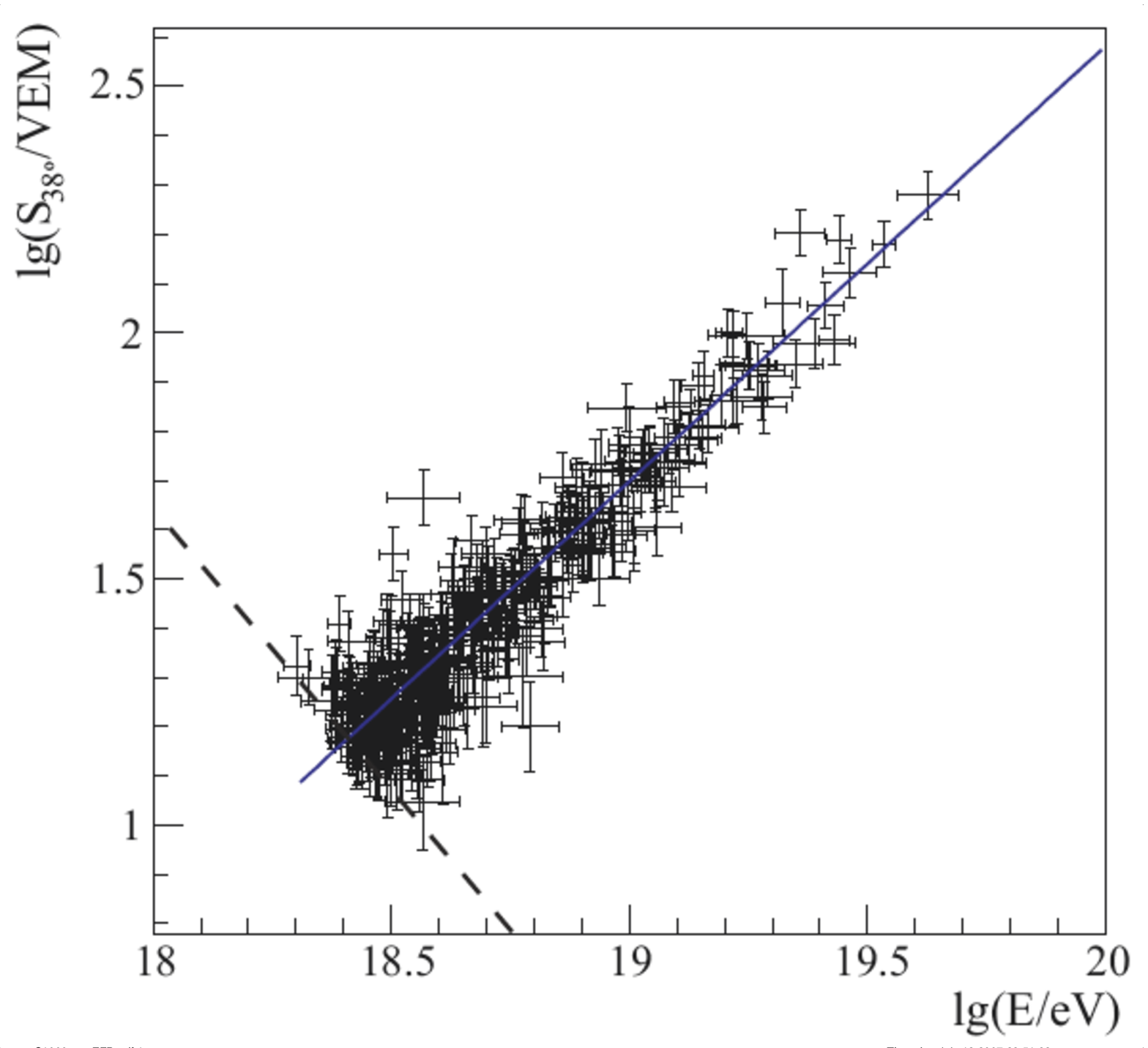}\hfill
\includegraphics[width=0.5\textwidth]{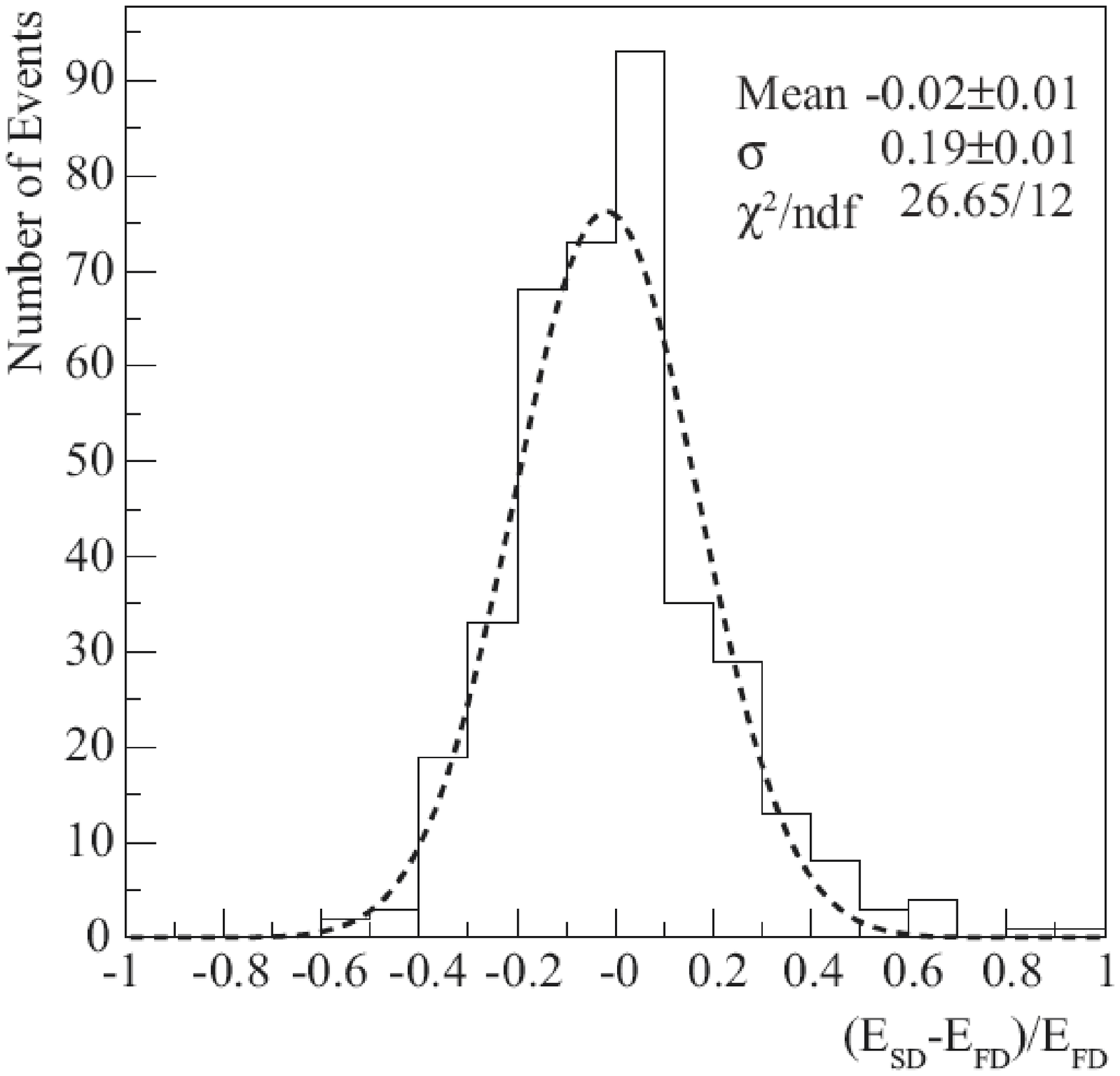}
\end{center}
\vspace{-1pc}
\caption{Left: correlation between the SD parameter $S_{38}$ (see text) and the reconstructed FD energy of Auger hybrid showers. Right: fractional dispersion of the SD/FD energy correlation (after iteration of the calibration procedure, see text).}
\label{fig:S1000_vs_EFD}
\end{figure}

Taking advantage of both aspects of the detector, the Auger Collaboration developed a cross-calibration technique enabling one to use the SD statistics with the FD energy scale measurement. As mentioned above, the intensity of the signal measured in a tank located 1000 meters away from the shower axis, noted $S_{1000}$, is directly related to the energy $E$ of the incoming CR. It can thus be used as an energy estimator, independently of shower simulations, provided one can experimentally establish a quantitative relation between $E$ and $S_{1000}$. This is done by systematically comparing the value of $S_{1000}$ and the energy $E_{\mathrm{FD}}$ reconstructed by the FD whenever it is available, i.e.~for all hybrid events passing the appropriate quality cuts. The result is shown in Fig.~\ref{fig:S1000_vs_EFD}a, exhibiting an excellent correlation, and the corresponding dispersion of the values is displayed on Fig.~\ref{fig:S1000_vs_EFD}b. The small relative dispersion (which includes uncertainties in the determination of both the FD energy and the SD signal) demonstrates that $S_{1000}$ is intrinsically a very good energy estimator, to provide reliable energy measurements once properly calibrated.

More precisely, the quantity that is plotted against $E_{\mathrm{FD}}$ in Fig.~\ref{fig:S1000_vs_EFD}a is not $S_{1000}$, but a modified quantity, $S_{1000}(38^\circ)$ (or $S_{38}$ for short), representing the $S_{1000}$ signal that would have been measured, had the shower developed at a zenith angle $\theta = 38^\circ$ (which happens to be the median of the Auger data set). The reason for using this modified quantity is that showers with the same energy developing at different zenith angles produce different $S_{1000}$ signals \emph{at ground level}, because the corresponding grammage of atmosphere along the shower axis (and thus the shower development stage, or ``age'') is different. Fortunately, it is in principle easy to relate $S_{1000}(\theta)$ to $S_{1000}(38^\circ)$, using the approximate isotropy of the observed CR flux: the value of $S_{1000}(\theta)$ (i.e.~$S_{1000}$ for a shower observed at zenith angle $\theta$) that corresponds to the same energy as a given value, $S_{38}$, measured for a shower at zenith angle $\theta = 38^\circ$, is the very value that gives the same integral flux of more-energetic CRs detected in the respective angular bin: $\int_{S_{1000}(\theta)}^\infty \Phi_{\mathrm{CR}}(S_{1000}(\theta))\mathrm{d}(S_{1000}(\theta)) = \int_{S_{1000}(38^\circ)}^\infty \Phi_{\mathrm{CR}}(S_{1000}(38^\circ))\mathrm{d}(S_{1000}(38^\circ))$, as determined experimentally by mere event counting, 

\begin{figure}[t]
\begin{center}
\includegraphics[width=0.5\textwidth]{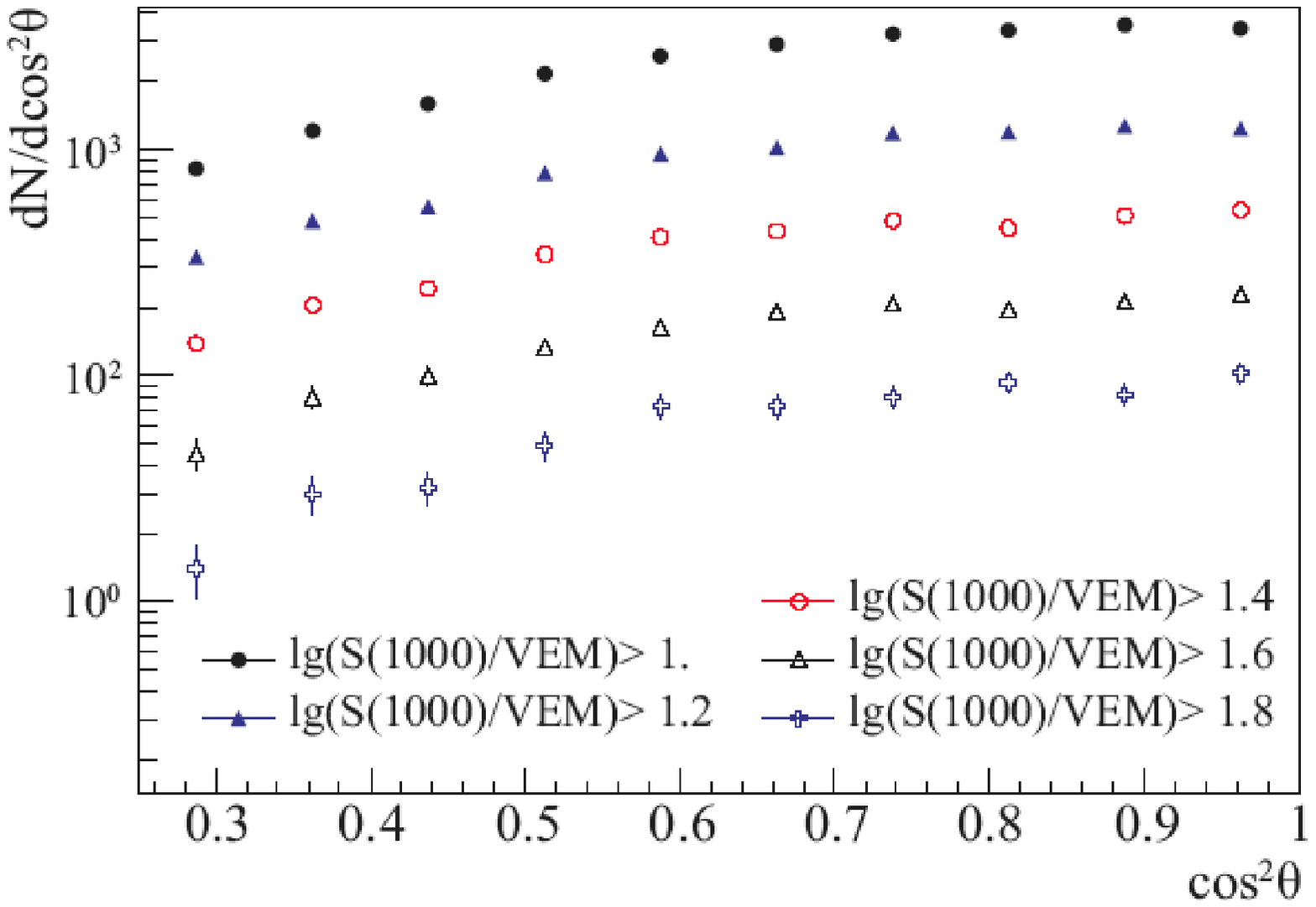}\hfill
\includegraphics[width=0.5\textwidth]{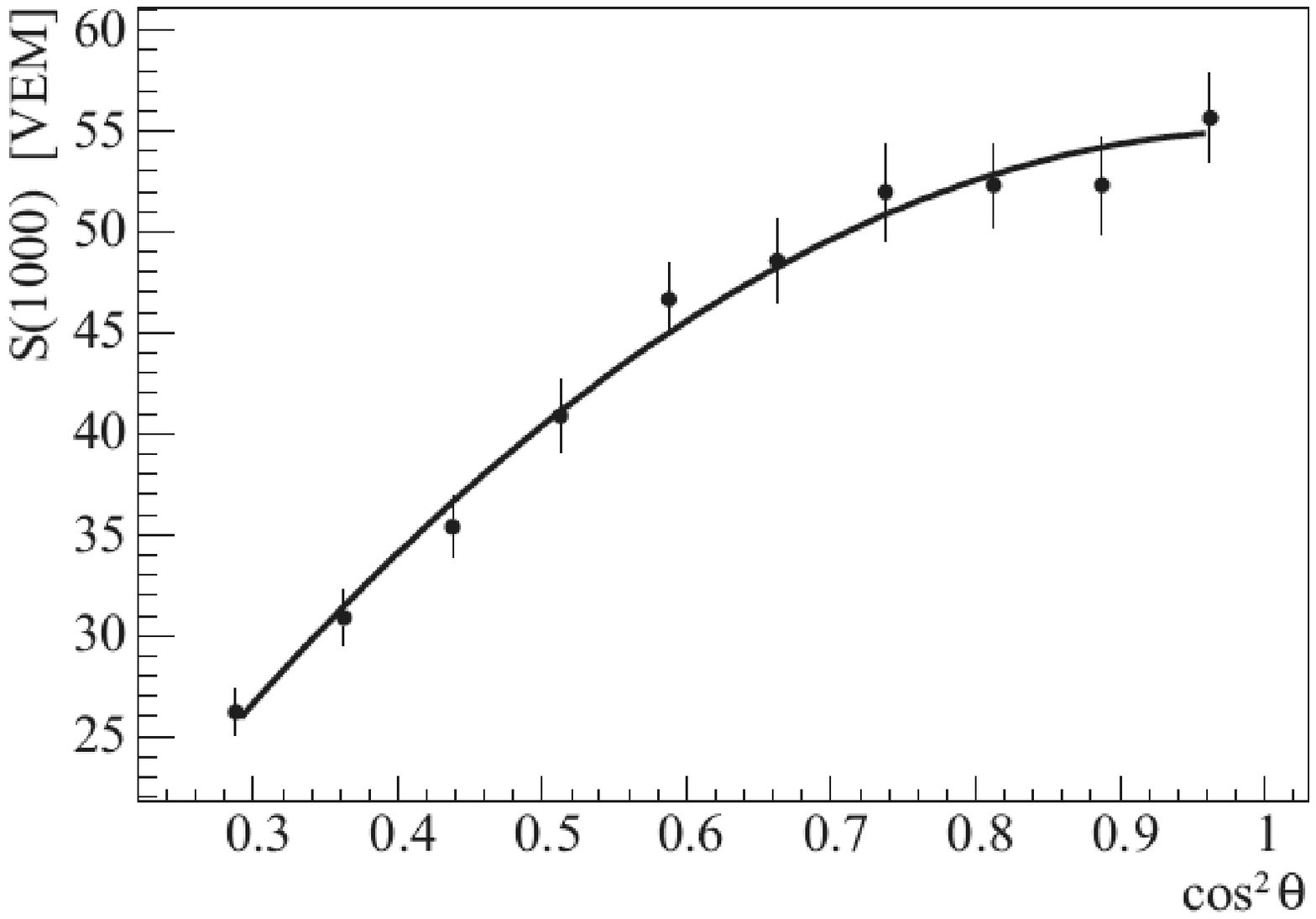}
\end{center}
\vspace{-1pc}
\caption{Left: total number of SD events above given values (indicated in ``VEM'', for ``vertical equivalent muons'') of the measured signal $S_{1000}(\theta)$, in different angular bins. Right: corresponding empirical attenuation curve, with a quadratic fit.}
\label{fig:CIC}
\end{figure}

This is made more explicit in Fig.~\ref{fig:CIC}a, where we plot the integral number of events in the Auger data set above five (arbitrarily chosen) values of $S_{1000}(\theta)$ spanning almost an order of magnitude in energy, in different zenith angle bins, regularly spaced in $\cos^2\theta$. As expected, this number is smaller at larger zenith angles (lower values of $\cos^2\theta$), since the showers are more attenuated and the same $S_{1000}$ signal corresponds to a larger energy, where the CR flux is smaller. Conversely, one can derive the so-called \emph{attenuation curve}, showing the evolution of $S_{1000}$ as a function of zenith angle for a given shower energy, i.e.~a given integral flux as in Fig.~\ref{fig:CIC}a. The result is shown in Fig.~\ref{fig:CIC}b, where the expected attenuation at large zenith angles is clearly seen. The line shows the empirical fit used to convert the value $S_{1000}(\theta)$ measured at zenith angle $\theta$ into the reference value $S_{38}$ corresponding to the same energy. The latter can then be considered as \emph{the} SD energy parameter, to be calibrated against measured FD energies for hybrid events, as already shown in Fig.~\ref{fig:S1000_vs_EFD}

Note that, in principle, the attenuation curve may vary with energy, since in addition to the attenuation itself (by which the signal is essentially reduced exponentially by the atmosphere grammage beyond the shower maximum), showers with higher energies develop lower in the atmosphere, and the most energetic ones may not even be fully developed when they reach the ground at low zenith angles. This effect was found to be negligible below a few $10^{19}$~eV, and the systematic uncertainty resulting from the assumption of a constant attenuation curve at higher energies was estimated conservatively. Obviously, these uncertainties will be reduced when the hybrid statistics keeps increasing and empirical attenuation curves can be derived at higher and higher energies.

To summarize, the energy of an SD event observed with zenith angle $\theta$ is derived in three steps:
\begin{enumerate}
\item determination of the signal 1000~m away from the shower axis ;
\item conversion into the signal that would have been measured at $38^\circ$ zenith angle (cf. Fig.~\ref{fig:CIC}): specifically, $S_{38} = S_{1000}/(1+ax + bx^2)$, where $a = 0.94\pm 0.06$, $b = -1.21\pm 0.27$ and $x = \cos^2\theta - \cos^2 38^\circ$~\cite{Roth_ICRC} ;
\item conversion into an FD-equivalent energy (cf. Fig.~\ref{fig:S1000_vs_EFD}): specifically, $\log E_{\mathrm{FD}} = A + B\log(S_{38})$, where $A = 17.08 \pm 0.03$ and $B = 1.13 \pm 0.02$~\cite{Roth_ICRC}.
\end{enumerate}

Finally, we note that promising methods to disentangle and take advantage of the hadronic and electromagnetic parts of the SD signal are being developed within the Auger Collaboration. Likewise, more precise measurements of the fluorescence yield will soon be available~\cite{Airfly}, helping reduce the uncertainties attached to the determination of the absolute energy scale. At present, the statistical and systematic uncertainties in the energy scale are $\sim 6$\% and $\sim 22$\%, respectively, the main contributions coming from the determination of the fluorescence yield (14\%), the absolute calibration of the FD (9.5\%) and the reconstruction method itself (10\%).

\section{A few chosen results}

The cosmic rays are characterized experimentally by three complementary spectral dimensions: the energy spectrum (differential flux), the angular spectrum (arrival directions), and the mass spectrum (composition). We briefly present here a few results related to these three dimensions, and refer the reader to the recent series of Auger publications presented at the 30th ICRC in July 2007 for further details and additional results.

\vspace{-12pt}
\subsection{Energy spectrum}

\begin{figure}[t]
\begin{center}
\includegraphics[width=0.5\textwidth,height=4.8cm]{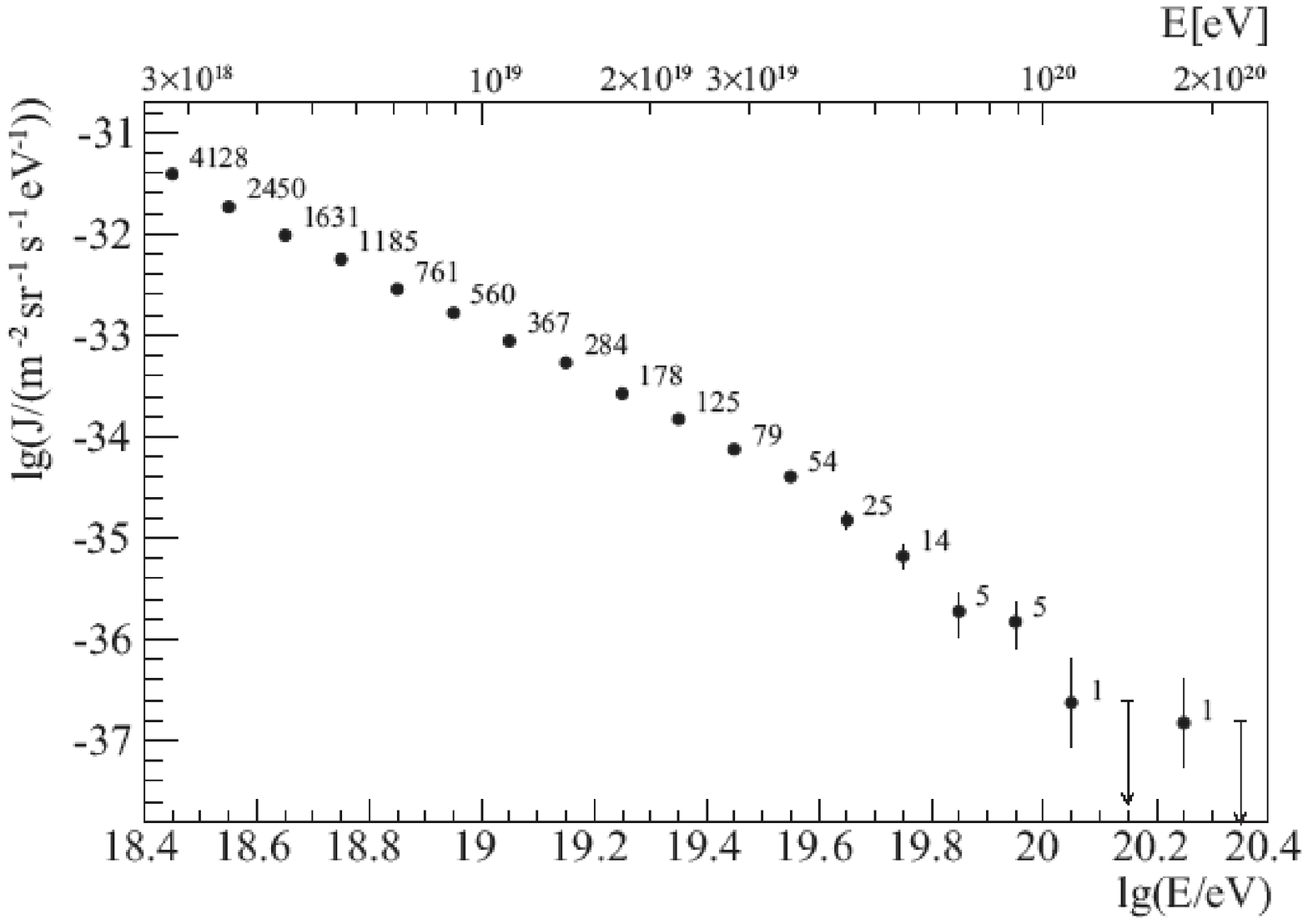}\hfill
\includegraphics[width=0.5\textwidth]{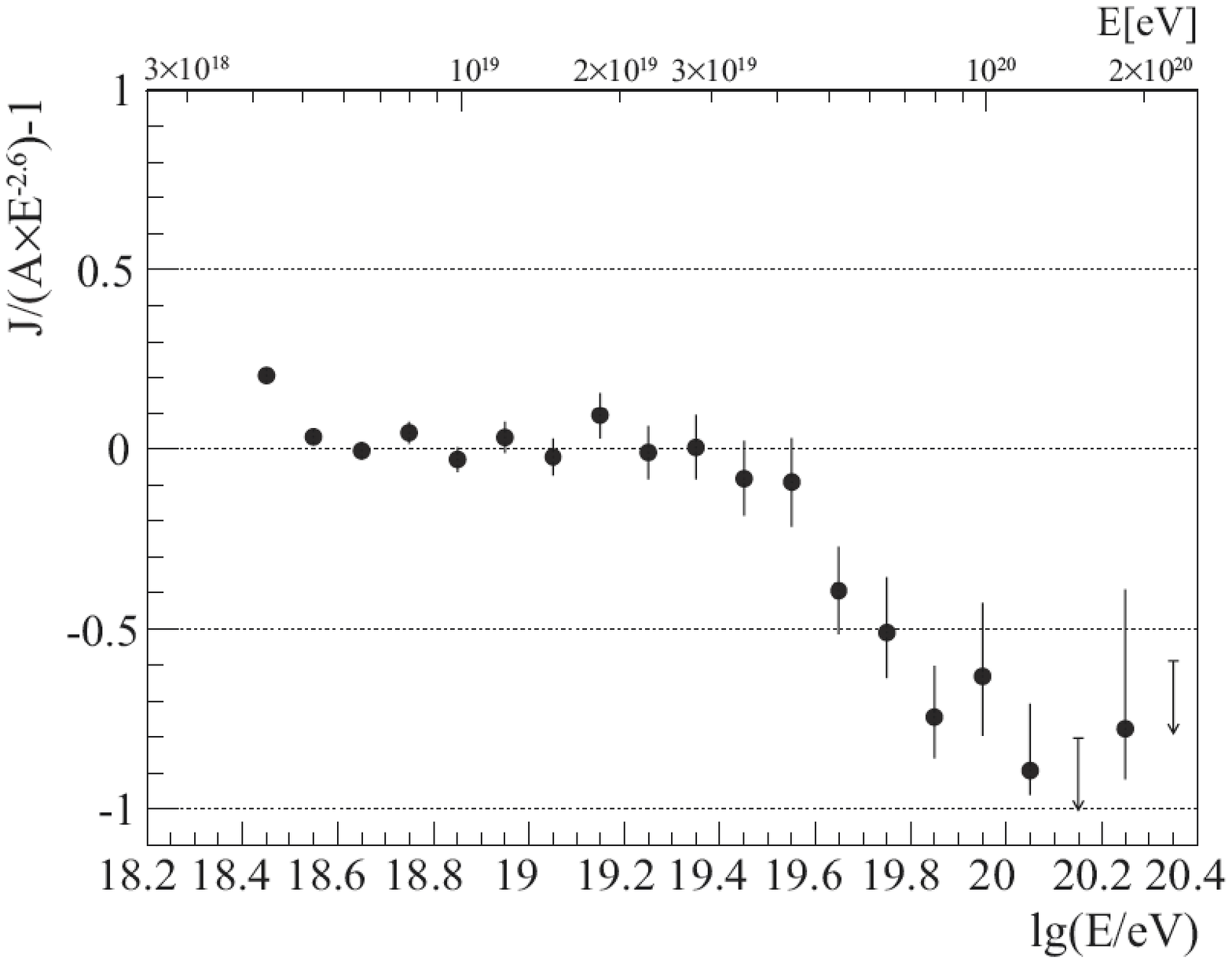}
\end{center}
\vspace{-1pc}
\caption{Left: Auger energy spectrum, with statistical error bars only (the number of events in each bin is indicated). Right: fractional difference between the Auger spectrum and an assumed CR flux in $E^{-2.6}$, as a function of energy.}
\label{fig:AugerSpectrum2007}
\end{figure}

The above-mentioned method was used to build the energy spectrum of cosmic rays above 3~EeV, with the full statistical power and controlled acceptance of the SD and the energy scale derived from the FD measurements. The result is shown in Fig.~\ref{fig:AugerSpectrum2007}a, where the number of events in each energy bin is indicated. The plot uses all the data gathered from the 1st of January, 2004, to the 28th of February, 2007, corresponding to an integrated exposure of 5165~km$^2$~sr~yr. In Fig.~\ref{fig:AugerSpectrum2007}b, we show the fractional difference between the Auger spectrum and a power-law in $E^{-2.6}$, which roughly corresponds to the measured spectrum between $10^{18.6}$~eV and $10^{19.6}$~eV (the actual best fit gives a logarithmic slope of $2.62\pm0.16^{[\mathrm{stat}]}\pm0.02^{[\mathrm{sys}]}$). A break is clearly seen: under the assumption of a continued power-law, one would expect $132\pm9$ events and $30\pm2.5$ events above $10^{19.6}$~eV and $10^{20}$~eV, respectively, while only 58~events and 2~events are observed. The hypothesis of a pure power-law can be rejected with a significance better than $6\sigma$ and $4\sigma$ for minimum energies of $10^{18.6}$ and $10^{19}$~eV, respectively\cite{Yamamoto_ICRC}.

\vspace{-12pt}
\subsection{Angular distribution}

Concerning the CR arrival directions, no significant departure from anisotropy can be reported yet. Previous tentative claims have been systematically studied, but the Auger data do not confirm them. This is the case in particular for the excesses from the Galactic center region claimed by the AGASA\cite{AGASAExcess} and SUGAR\cite{SUGARExcess} experiments\cite{AugerAnisotropies}. Large angular scale anisotropy were also searched for in different energy ranges. Several methods were used and their sensitivity was considerably improved thanks to a detailed study of various systematic effects related to the irregular growth of the SD array, the slight inhomogeneity of the tank response across the array and subtle weather effects involving both pressure and temperature. At present, the right-ascension (RA) distribution of the events appears remarkably isotropic at EeV energies, with an upper limit of 1.4\% on the first harmonic amplitude (dipole in RA modulation)\cite{Armengaud_ICRC}.

Searching for angular coincidences between Auger events and BL Lac objects, no significant correlation could be found.

Finally, searching for a signal related to the possible clustering of high-energy Auger events, no strong excess was observed, particularly not as strong as suggested by the AGASA results at a scale of 2.5$^\circ$ above 40~EeV\cite{AGASAClustering}. An extensive scan in angle and energy threshold showed some hints of clustering at larger energies ($E > 50$~EeV) and intermediate angular scales, which could reveal the large scale distribution of nearby sources. However, taking into account the scan performed, the overall chance probability of such a signal from an isotropic flux is only 2\%, i.e.~only marginally significant with the present statistics.

\vspace{-12pt}
\subsection{Composition}

\begin{figure}[t]
\begin{center}
\includegraphics[width=0.5\textwidth]{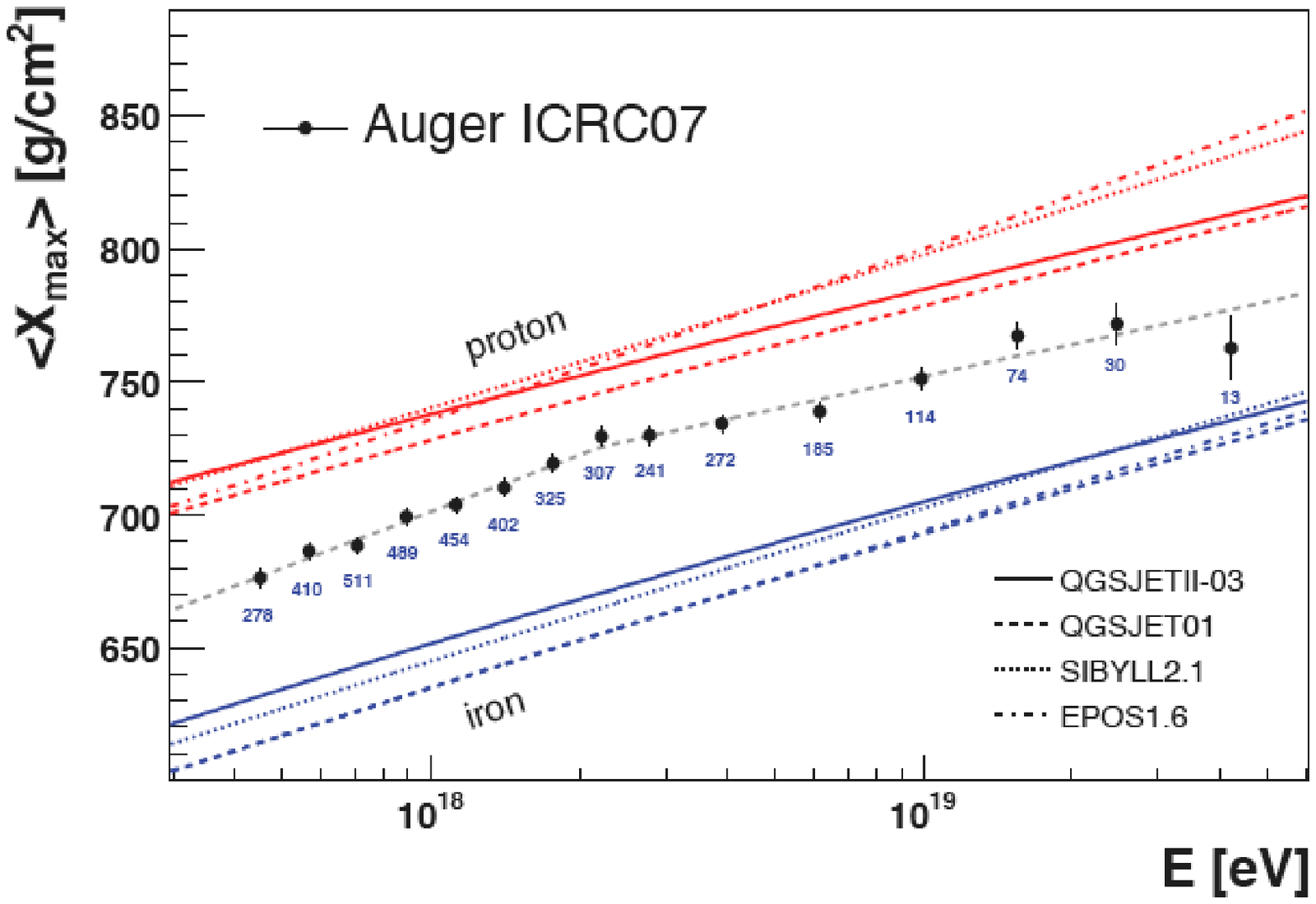}\hfill
\includegraphics[width=0.5\textwidth]{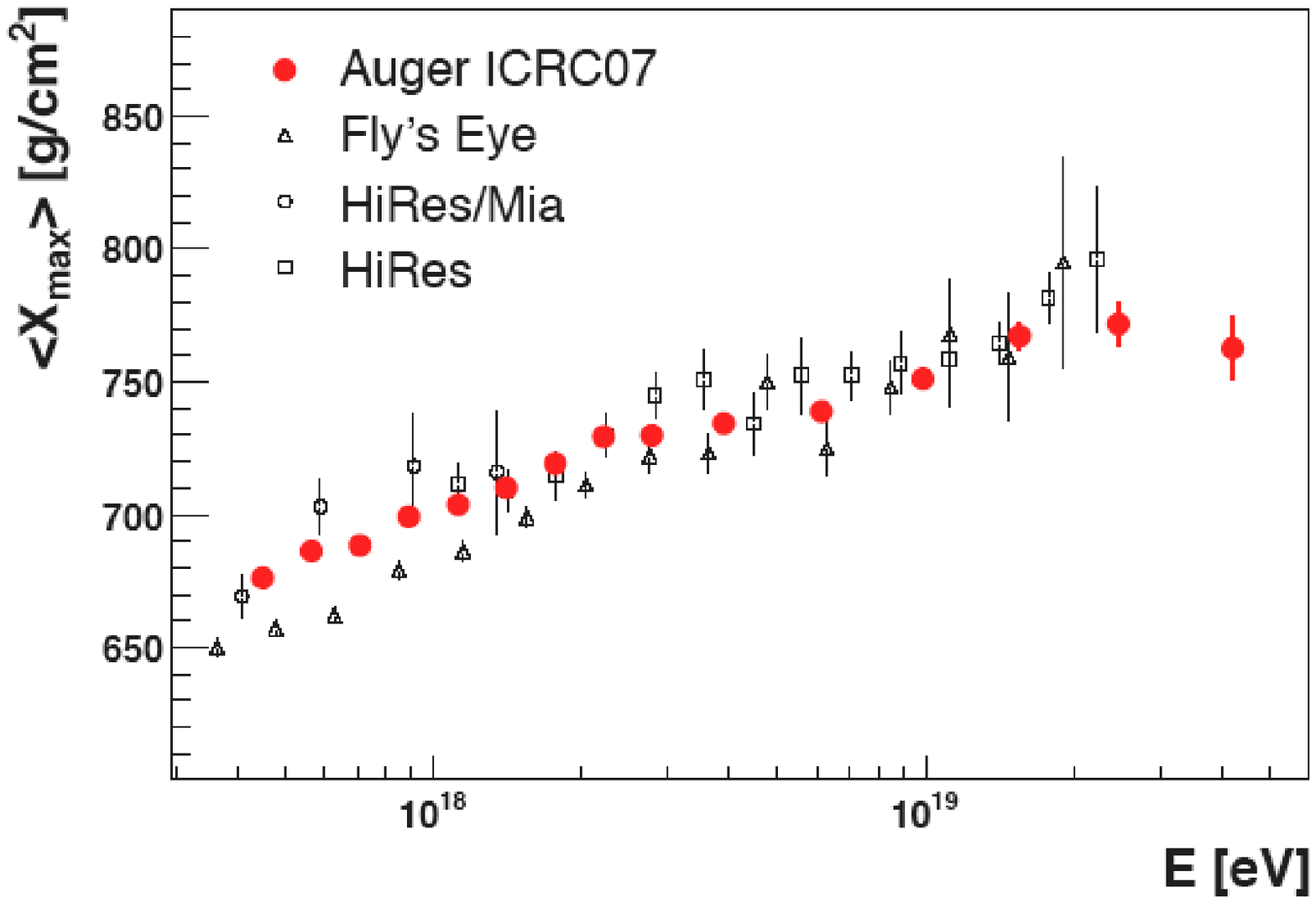}
\end{center}
\vspace{-1pc}
\caption{Left: evolution with energy of the average atmospheric grammage at shower maximum, $\left<X_{\max}\right>$. The number of hybrid events used in each energy bin is indicated, and the expectations for pure proton and pure Fe UHECRs from different hadronic interaction models are shown for reference. Right: same data, plotted together with the previous results of Fly's Eye and HiRes.}
\label{fig:AugerXMax2007}
\end{figure}

High-energy CR observatories do not detect the primary cosmic rays themselves, but the induced showers in the atmosphere. To identify the nature of the primaries, one must thus look for differences in the shower development, which are usually relatively small and subject to fluctuations associated with the stochasticity of the first interactions. Essentially, at a given energy, showers from heavier nuclei develop slightly earlier (or quicker) in the atmosphere, and give rise to smaller fluctuations. On average, they reach their maximum development higher in the atmosphere, i.e.~at a lower cumulated grammage, $X_{\max}$. In parallel, $X_{\max}$ is naturally increasing with energy, as more energetic showers can develop longer before being quenched by atmospheric losses. This is shown in Fig.~\ref{fig:AugerXMax2007}, where the upper and lower lines correspond to the expectations for the evolution of the average $\left<X_{\max}\right>$ with energy, for pure protons (upper lines) and pure Fe nuclei (lower lines), under different assumptions for the modeling of high-energy hadronic interactions. Typically, one may write: $\left<X_{\max}\right> = D_{p}[\ln(E/E_{0})-\left<\ln A\right>] + c_{p}$, where $\left<\ln A\right>$ is the average logarithmic mass of CRs at energy $E$, and $D_{p}$ and $c_{p}$ are constants that depend on the assumed hadronic model. Fig.~\ref{fig:AugerXMax2007}a shows the results obtained with a subset of the Auger hybrid data, satisfying appropriate quality and uniformity cuts, which guarantee an $X_{\max}$ resolution at the 20~g/cm$^2$ level\cite{Unger_ICRC}.

Several features can be observed: first, an evolution with a slope larger than that of either pure iron or pure proton compositions, up to 2--3~EeV, indicating a lightening of the primary CR composition ; then a flattening of the evolution, an inflection point and a steepening again up to 20--30~EeV (possibly indicating complex composition changes, as anticipated in~\cite{AllardEtAlCompo}) ; and finally a last point showing a decrease of $\left<X_{\max}\right>$, indicating a change to a heavier composition. Although a few points at higher energy would be needed to strengthen this result, it seems to indicate that the UHECRs are not made of protons only (unless some exotic or unexpected hadronic physics can explain the behaviour of $\left<X_{\max}\right>$ above 10~EeV).

\begin{figure}[t]
\begin{center}
\includegraphics[width=0.5\textwidth,height=5cm]{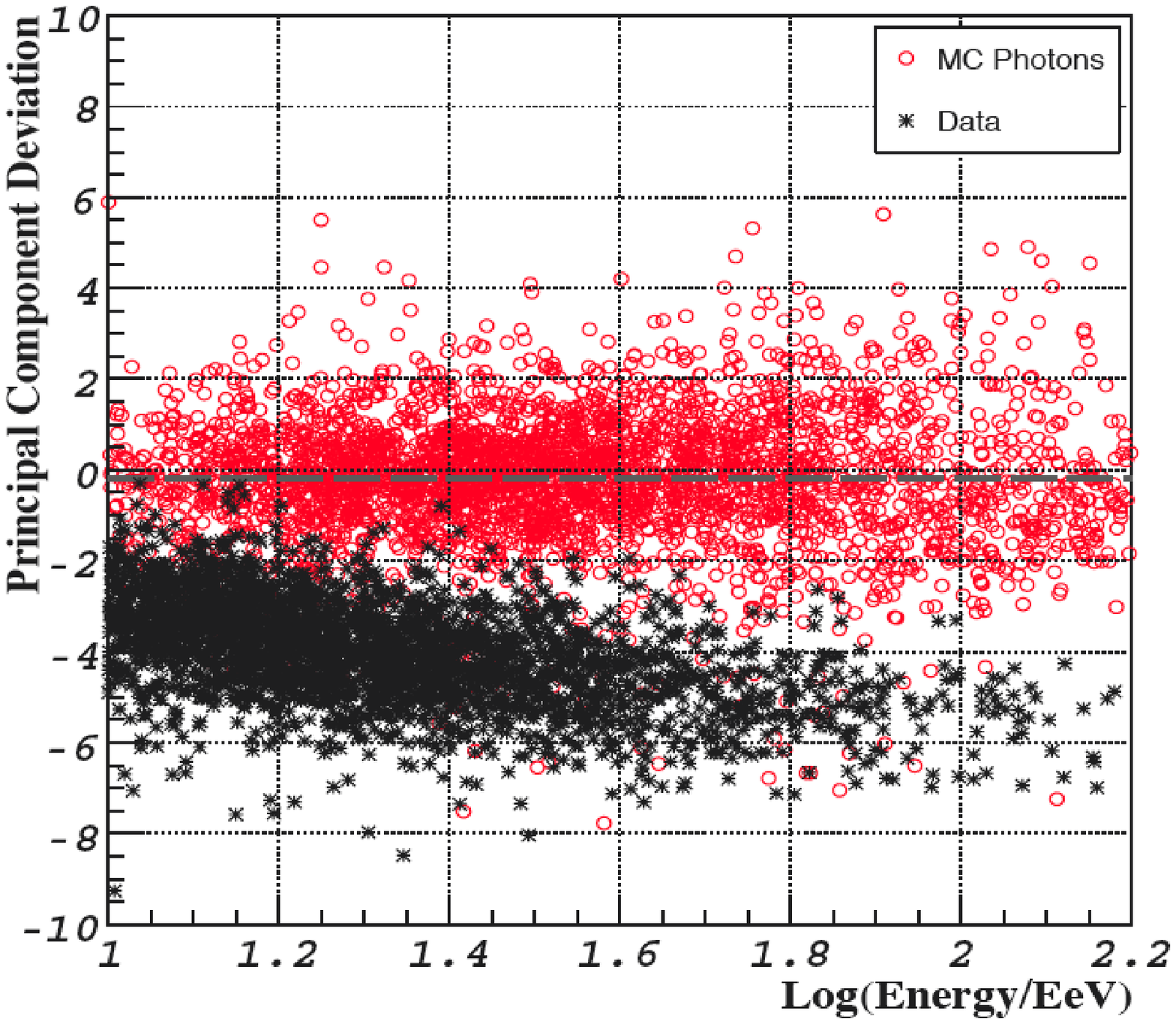}\hfill
\includegraphics[width=0.5\textwidth,height=5cm]{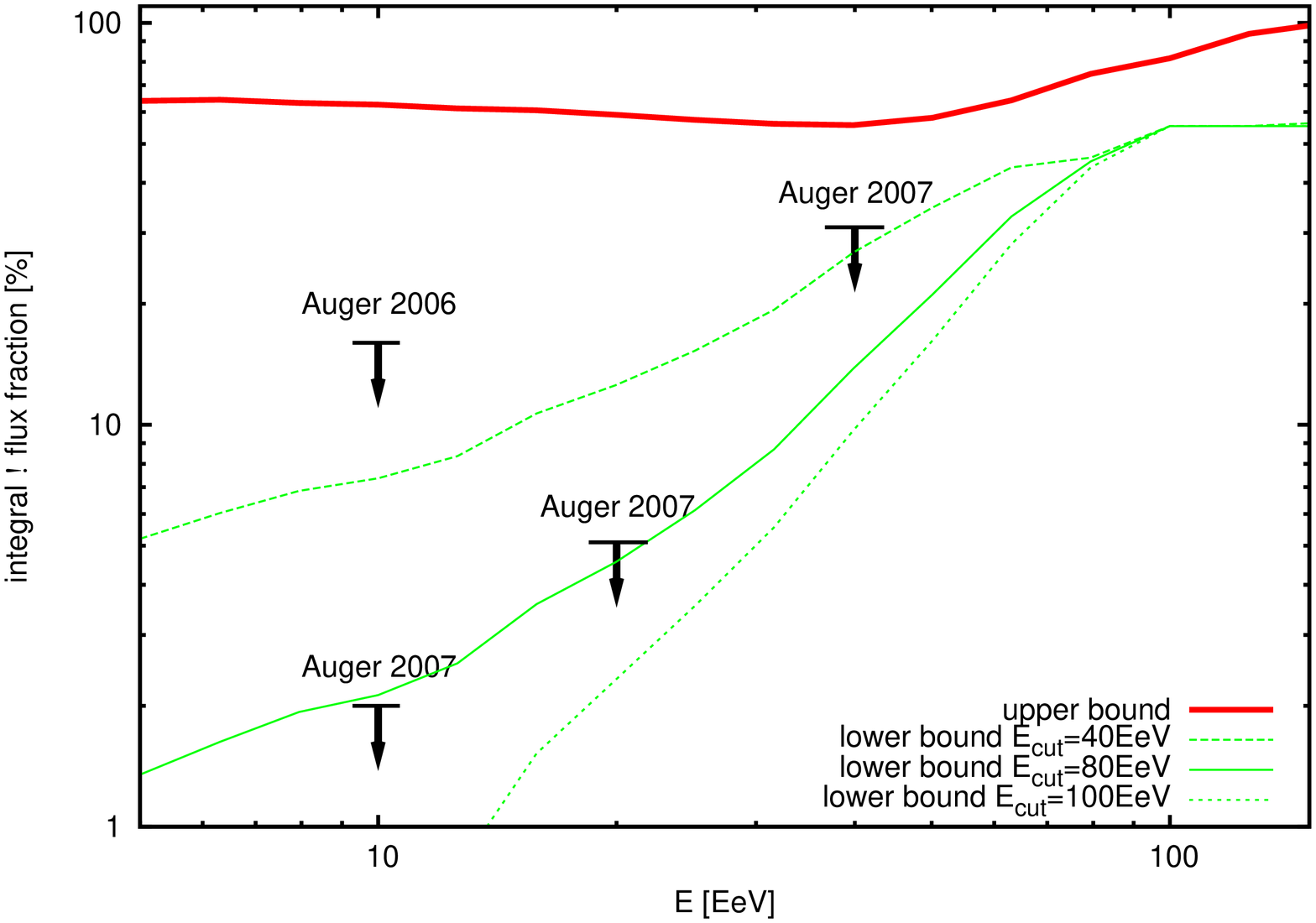}
\end{center}
\vspace{-1pc}
\caption{Left: principal component between rise time and radius of curvature for simulated photon-induced showers (red circles) and for the SD data (black dots), as a function of energy. Right: photon fraction limit among UHECRs, compared with SHDM models predictions (see text).}
\label{fig:AugerPhotonLimit}
\end{figure}

Primary photons are easier to distinguish from nuclei than nuclei between themselves. The main reason is that the particle-producing reactions involved in the development of electromagnetic-dominated photon-induced showers have a much smaller multiplicity than in hadronic showers. The photon showers therefore develop more slowly and penetrate more deeply into the atmosphere (and even more so above 30~EeV, where the suppression of the Bethe-Heitler pair production due to the LPM effect becomes important\cite{LPMEffect}). The result is a much larger $X_{\max}$, which provides an efficient discrimination tool for hybrid events. In turn, the larger $X_{\max}$ results in a larger rise time of the signal produced in the SD tanks (defined as the time needed to pass from 10\% to 50\% of the total signal), and in a smaller radius of curvature of the shower front. These two parameters can be measured accurately for all SD events, and combined into a single SD observable through the principal component analysis to maximize the discrimination power. The distribution of this parameter for (simulated) photon showers is shown in Fig.~\ref{fig:AugerPhotonLimit}a, together with the measured values from the Auger data set. Any shower with a principal component above the mean of the photon distribution is considered a photon candidate. None has been found so far, at any energy, which leads to the upper limits shown in Fig.~\ref{fig:AugerPhotonLimit}b for the fraction of photons among high-energy CRs.

The absence (or low fraction) of photons at high-energy sets important constraints for a whole class of UHECR source models known as \emph{top-down models}. In such models, the UHECRs are decay products of high-mass particles (above $10^{21}$~eV) either inherited from the early universe and gathering in the halo of galaxies and clusters of galaxies (e.g. super-heavy dark matter, or SHDM models), or produced throughout the universe by the interaction and/or annihilation of topological defects (TD models). In all cases, the UHECRs are produced as secondary particles through a hadronization process and mostly made of photons and neutrinos, with a minority of nucleons. In the case of SHDM models, most UHECRs detected on Earth would come from our own Galactic halo, with no propagation effect, i.e.~we would observe essentially photons with a hard energy spectrum (which could thus contribute significantly to only the last few bins of the detected CR flux). The new photon limit obtained by Auger essentially excludes all such models, as shown in Fig.~\ref{fig:AugerPhotonLimit}b. In the case of TD models, the UHECR sources are distributed throughout the universe and most of the high-energy photons interact with the CMB before reaching the Earth, which suppresses the photon component relatively to the protons. The predicted photon fraction is thus smaller in this case, and some TD models can escape the Auger limit (see~\cite{Semikoz_ICRC} for more details). The neutrino flux, however, is not attenuated, and the predicted level for such models may also be challenged by the Auger neutrino limit in the future.

\section{Summary and perspective}

The Southern site of the Pierre Auger Observatory has now been taking data with increasing acceptance for more than three years, accumulating a total exposure of more than 5000~km$^2$~sr~yr. Deployment will be over within a year. We have described the method currently used to analyze the Auger data: it takes advantage of the hybrid nature (SD and FD) of the detector, cross-calibrating the SD and the FD to exploit the large statistics collected by the SD while keeping an essentially hadronic-model independent measurement of the energy.

A few results have been presented. Thanks to the large statistics available, a sharp suppression of the UHECRs in the last decade of the observed energies (as already shown by the HiRes experiment~\cite{HiResAbbasiEtAl2005}) could be established with a high confidence level. Whether this is due to the expected GZK effect or to a limit of the acceleration process is still an open question. A definitive proof of the GZK effect, best captured as a reduction of the source horizon at high energy, would be the detection of a correlation between the highest energy events and nearby sources (i.e.~with an upper limit on redshift). This is expected to be observed relatively soon if the extragalactic magnetic fields are not significantly larger than anticipated from current measurements and simulations.

This would mark the opening of the so-called \emph{proton astronomy}, where some global patterns will first be observed in the sky and eventually individual sources can be distinguished. This will of course allow considerable progress in the identification of the responsible cosmic accelerators, especially when the integrated exposure is sufficient to draw individual source spectra. Other progress should accompany such developments, as the UHECRs can then be used as diversely as to probe Galactic and extragalactic magnetic fields, to shed light on the acceleration mechanims at work, to constrain the overall energy budget at the source, and more generally as complementary messengers from powerful sources in the universe (together with radio waves, X-rays, gamma-rays and other thermal and non-thermal radiations, and possibly in the future with neutrinos and gravitational waves). The Southern site of the Pierre Auger Observatory should be the first to open this new era. However, the challenges of proton astronomy will soon require much larger exposures, as well as a full coverage of the sky. The next obvious step in this direction appears to be \emph{Auger North}, i.e.~the Northern site of the observatory to be installed in Colorado, USA. Then considerably larger exposures may require experiments observing the cosmic-ray atmospheric showers from space, possibly at the price of increasing the energy threshold of the detectors. In any case, it is clear that decisive progress in the field of UHECRs will begin when one can draw spectra from individual sources or at least limited regions of the sky, rather than having to cope with the overall spectrum summing the contribution of all sources everywhere in the universe!

In the meantime, special attention should be devoted to lower energy CRs, in the EeV range, where a concave feature known as the ``ankle'' is observed in the energy spectrum. Two different interpretations have been proposed for the ankle: it may be a spectral feature carved by pair-production energy losses suffered by UHECR protons while they propagate through the CMB\cite{BerezinskyEtAlPPDip}, or the sign of a gradual transition from a steep Galactic CR component to a flatter extragalactic component\cite{AnkleTransitionAllardEtAl}\cite{AllardEtAlCompo}. These interpretations are associated with a very different UHECR phenomenology, notably concerning the CR source composition (pure or almost pure protons in the first case, standard -- i.e.~similar to low-energy CRs -- mixed composition in the second case) and UHECR source spectrum (steeper in the first case, say $\propto E^{-2.6}$, than in the second, say $\propto E^{-2.3}$)\cite{AnkleTransitionAllardEtAl}. Furthermore, since the Galactic/extragalactic transition does not occur at the same energy in both cases, a better understanding of the ankle will have implications on low-energy Galactic cosmic rays as well, whose sources are still unknown, as should always be remembered when dealing with CR-related issues.

While the ankle may be a key to understanding the global CR phenomenon, stronger constraints on the CR composition and spectrum in the EeV range may be the key to understanding the ankle. To this end, the Auger collaboration is now developing ``detector enhancements'', notably through two complementary experiments to be installed on the site of Auger and fully integrated to its operation: HEAT, consisting of High Elevation Auger Telescopes identical to the Auger FD telescopes but tilted upwards by 30$^{\circ}$, to observe showers down to 0.1~EeV and measure their $X_{\max}$ higher in the atmosphere\cite{Klages_ICRC}), and the associated AMIGA, a 23.5~km$^2$ array of SD detectors deployed on a denser grid with spacings of 433~m and 750~m, complemented by 30~m$^2$ burried muon scintillator counters to detect showers down to 0.1~EeV and measure their muon content, a powerful handle to the CR composition\cite{Etchegoyen_ICRC}.

More results and further details about the detector's design, performance, operation and achievements can be found in other Auger publications (see notably the series of papers presented at the 30th ICRC).

\end{document}